\documentclass[12pt,a4paper]{article}

\usepackage[textheight=23cm,textwidth=15cm]{geometry}
\usepackage{amsmath}
\usepackage{graphicx}
\usepackage[american]{babel}
\usepackage{here}
\usepackage[articletitle=true, super=true, maxauthors=1000]{achemso}
\setcitestyle{square}
\usepackage{multirow}

\usepackage[hidelinks]{hyperref}

\usepackage{url}
\usepackage{xcolor}

\begin{document}

\renewcommand{\labelitemii}{$\longrightarrow$}

\begin{center}

{\LARGE\bf
  Automated construction of\\[0.3em] quantum--classical hybrid models
}

\vspace{1cm}

\renewcommand*{\thefootnote}{\fnsymbol{footnote}}

{\large
Christoph Brunken
and
Markus Reiher\footnote{Corresponding author; e-mail: markus.reiher@phys.chem.ethz.ch}
}\\[4ex]

\renewcommand*{\thefootnote}{\arabic{footnote}}
\setcounter{footnote}{0}

ETH Z\"urich, Laboratorium f\"ur Physikalische Chemie, Vladimir-Prelog-Weg 2,\\ 
8093 Z\"urich, Switzerland \\

\vspace{.5cm}

February 18, 2021

\vspace{.43cm}

\textbf{Abstract}
\end{center}
\vspace*{-.41cm}
{\small
We present a protocol for the fully automated construction of quantum mechanical-(QM)--classical hybrid models by extending our previously
reported approach on self-para\-metri\-zing system-focused atomistic models (SFAM) [\textit{J. Chem. Theory Comput.} \textbf{2020}, \textit{16}(3), 1646--1665].
In this QM/SFAM approach, the size and composition of the QM region is evaluated
in an automated manner based on first principles so that the hybrid model describes the atomic forces in the center of the QM region accurately.
This entails the automated construction and evaluation of differently sized QM regions with a bearable
computational overhead that needs to be paid for automated validation procedures.
Applying SFAM for the classical part of the model eliminates any dependence on pre-existing parameters due to its system-focused quantum mechanically
derived parametrization. Hence, QM/SFAM is capable of delivering a high fidelity and complete automation.
Furthermore, since SFAM parameters are generated for the whole system, our ansatz allows for a convenient re-definition of the QM region
during a molecular exploration. For this purpose, a local re-parametrization scheme is introduced, which efficiently generates additional classical parameters on the
fly when new covalent bonds are formed (or broken) and moved to the classical region.
}

\newpage


\section{Introduction}
\label{sec:introduction}

In contrast to most protocols of computational quantum chemistry that consider isolated molecules,
chemical processes can take place in a vast variety of complex environments.
Studying chemical reactions in proteins, in nanostructures, and on surfaces, requires a theoretical approach that must provide an accurate quantum mechanical description
of the reaction center and at the same time an efficient model to cope with the enormous system size of the structured, heterogeneous environment.
For these requirements to coalesce, one is forced to apply a hybrid method, which divides the
system into several regions treated at different levels of approximation.\cite{senn06, groenhof13}

Typically, the reaction center is modeled with a quantum mechanical method, which allows one to describe the formation and cleavage of covalent bonds in a natural way.
The environment may efficiently be treated by a force field~\cite{riniker18} rooted in classical mechanics.
Such a quantum-mechanical/molecular-mechanical hybrid (QM/MM) model~\cite{warshel76, thiel13} typically requires
significant manual work involved in the model construction process. It is plagued by a lack of standardization and comparability,
and rigorous uncertainty quantification is not available (as for most computational chemistry methods), which is particularly critical for such complex composite models.
Typically, no standardized procedures exist regarding, for instance, the parametrization of the force field (especially for metal-containing regions),
the QM region determination, the choice of boundary scheme, the initial structure generation, conformational sampling, and the extent
to which parts of the macromolecule are constrained during structure optimizations in order to confine the complexity of the system as well as to limit the error of the MM energy contribution.

One of the most prominent challenges for QM/MM modeling is the construction of an efficient and at the same time accurate molecular mechanics model, which is typically only available for a
pre-defined subset of chemical elements with standard bonding patterns. Its degree of transferability to a new system is not obvious at all.
We addressed this issue in our recent work~\cite{brunken20} by introducing the self-parametrizing system-focused atomistic model~(SFAM), which allows for an automated
construction of a molecular mechanics model for a system of arbitrary size and elemental composition. The reference data for the model are obtained in a fully automated way
by an autonomous fragmentation algorithm and subsequent
quantum chemical reference calculations for the molecular fragments. Furthermore, SFAM includes a model refinement step
based on $\Delta$-machine learning~\cite{ramakrishnan15} when additional reference data become available during a molecular exploration.

In this work, we extend SFAM toward quantum--classical hybrid models (QM/SFAM) with an automated set-up.
The application of SFAM as the classical part of the model eliminates any issues arising from an incomplete set of parameters.
Furthermore, SFAM guarantees that the reference data, from which the MM parameters are derived,
can be provided by the same quantum chemical method as is applied in the QM part of the hybrid calculation
making the MM model as consistent as possible with the QM part.

Furthermore, we present a scheme to determine the choice of the QM region which results in an accurate QM/SFAM model.
The selection of atoms to include in the quantum mechanical part of the calculation is a highly non-trivial task and the decision often relies on chemical intuition alone. However, there have been recent efforts
to select QM regions systematically on a first-principles basis by Kulik and coworkers~\cite{karelina17}. Their systematic QM region determination scheme relies on evaluating how specific residues of a protein
affect the electronic structure (charge distribution, frontier orbitals) of the core residues of the protein. In this work, we provide an alternative approach solely based on the fundamental quantities of a molecular system,
i.e., its electronic energy and its derivatives with respect to the nuclear coordinates, which are available from 
computationally cheap and approximate as well as expensive and accurate electronic structure models. 

The nuclear derivatives play an essential role in obtaining reliable structures (by molecular dynamics (MD)
sampling or structure optimization)
and we therefore focus on the accurate description
of the atomic forces when determining which atoms must be included in the quantum mechanical part of the model.
Defining an optimal set of atoms as the QM region is essential, especially because
it has been demonstrated by Ochsenfeld~\cite{sumowski09, flaig12, rossbach17}, Martinez~\cite{kulik16} and others~\cite{hu11, liao13, retegan13} that for many systems
QM/MM models safely converge only with large QM region sizes of several hundred atoms. Therefore, in those cases where such large QM regions are not feasible
(e.g., vast reaction network explorations or MD simulations), it is inevitable to carefully select the QM region systematically in order to guarantee that the resulting model
is an accurate approximation to a full quantum mechanical model.

In the following, we first describe our QM/SFAM model and then introduce an algorithm to systematically determine the composition of the QM region in an automated way.
This is demonstrated with the examples of (i) a medium-sized peptide
that also allows for full-quantum reference calculations and of (ii) a larger system to resemble a typical case of application.
Although these examples are taken from biochemistry, we emphasize that our model is agnostic with respect to the elemental
composition due to its first-principles core. Hence, any nanoscale atomistic system can be subjected to our hybrid model
construction process, even one for which a molecular structure first needs to be constructed (by virtue of the
SFAM approach that early on in the model generation provides an approximate force field for iterative structure refinement \cite{brunken20}).

\section{Theory}
\label{sec:theory}

\subsection{The SFAM approach}
\label{sec:sfam_approach}

We briefly review the SFAM approach~\cite{brunken20} as it will be the classical part of our hybrid model. 
Similar to QMDFF~\cite{grimme14} and QuickFF~\cite{vanduyfhuys15} molecular mechanics models, also SFAM is
generated automatically for a specific molecular system from quantum mechanical reference data, which
yields accurate force fields without being limited by the elemental composition of the molecular system.
SFAM is distinct from the two aforementioned models~\cite{grimme14, vanduyfhuys15} in two crucial aspects.
First, SFAM force constants are parametrized by a partial Hessian fit algorithm~\cite{wang16, wang18} as introduced by Hirao and coworkers in 2016, i.e., the parameters are fitted solely to local information in the Hessian,
which allows us to generate the model for very large molecular systems by calculating reference data for fragments cut out of the whole structure.
We also introduced an autonomous fragmentation algorithm for this purpose.~\cite{brunken20}
Second, SFAM includes an (optional) improvement step based on $\Delta$-machine learning~($\Delta$-ML)~\cite{ramakrishnan15}. 

While the MM base model of SFAM provides an accurate
description of the potential energy surface~(PES) close to
the local energy minimum taken as a reference for parametrization, its parameters are not guaranteed to be transferable across all regions of the PES. The base model can be applied in an exploration of additional structures
(e.g., in molecular dynamics simulations), for which additional reference data can be calculated on the fly. The MM/ML ansatz of SFAM 
can then gradually increase its accuracy across the PES as an
increasing amount of reference data is collected to train the ML model.
Many ML-only models have been reported as replacements for classical force fields~\cite{li15, chmiela17, glielmo17, chmiela18, amabilino20}.
However, our MM/ML approach for SFAM has several advantages.
On the one hand, the MM base model provides physical insight into the properties of the system in contrast to an approach solely based on ML. On the other hand, it
requires only a limited and well controllable amount of reference data, as it is parametrized on
single-point data obtained for fragments (optimized structures, atomic charges and Hessians).

The SFAM energy $E_\text{SFAM}$ can be written as the sum of the MM and ML contributions,
\begin{equation}
 E_\text{SFAM} = E_\text{MM} + E_\text{ML} \quad .
\end{equation}
$E_\text{ML}$ is zero at this level, because we choose 
the hybrid QM/SFAM model as the new base model, which can then be refined in a later step by $\Delta$-ML.
The SFAM energy expression is divided into a covalent (cov)
part $E_\text{cov}$ and nonbonding (nb) potential energy contributions $E_\text{nb}$,
\begin{equation}
 E_\text{SFAM} = E_\text{cov} + E_\text{nb} \quad ,
\end{equation}
as is common in MM models~\cite{riniker18}. The covalent energy contribution is calculated from the displacements of the internal degrees of freedom out of their equilibrium positions and can therefore be
divided into terms for bonds $r$, bond angles $\alpha$, dihedral angles $\theta$, and improper dihedral angles $\varphi$,
\begin{align}
 E_\text{cov} &= E_r + E_\alpha + E_\theta + E_\varphi \nonumber \\ 
 &= \sum_{(A,B)} E^{AB}_r + \sum_{(A,B,C)} E^{ABC}_\alpha + \sum_{(A,B,C,D)} E^{ABCD}_\theta + \sum_{(X,B,C,D)} E^{XBCD}_\varphi \quad ,
\end{align}
where $(A,B)$ denotes a group of two bonded atoms $A$ and $B$, $(A,B,C)$ a bonded triplet of atoms, $(A,B,C,D)$ a bonded quadruplet of atoms, and $(X,B,C,D)$ the atoms of
an improper dihedral angle~(with $X$ representing the center atom).
The nonbonding interactions comprise an electrostatic part (estat), dispersive (disp) and Pauli repulsion (rep) interactions,
and hydrogen bonds,
\begin{align}
 E_\text{nb} &= E_\text{estat} + E_\text{disp} + E_\text{rep} + E_\text{hb} \nonumber \\ 
 &= \sum_{(A,B)} \left( E_\text{estat}^{AB} + E_\text{disp}^{AB} + E_\text{rep}^{AB} \right) + \sum_{(D,H,A)} E_\text{hb}^{DHA} \quad ,
\end{align}
where $(A,B)$ represents a pair of atoms, in which the atoms $A$ and $B$ are neither bonded to one another nor both bonded to another 
atom $C$, and $(D,H,A)$ is a hydrogen bond.
For details on the potential energy expressions for each of the MM contributions, as well as an explanation of the parametrization procedure, we refer to our previous paper.~\cite{brunken20}.

We emphasize that combining the classical model in SFAM with a quantum chemical method creates an opportunity to apply our $\Delta$-ML improvement step to the hybrid model.
During a molecular exploration with the QM/SFAM method, quantum chemical reference data are collected without any additional effort. 
Our machine-learned model corrections can be trained with these data and
will be valuable (i) if the QM focus is moved to a different section of the whole system
and (ii)  if the data can be transferred to improve the description of similar molecular substructures located in the MM region.
We also note that related efforts to combine machine learning with quantum$-$classical hybrid methods have been reported recently.~\cite{zhang18, boselt20}

\subsection{The hybrid QM/SFAM formalism}
\label{sec:qm_mm_formalism}

The energy expression of any hybrid model with two distinct regions,
a quantum core $\mathcal{Q}$ and an environment $\mathcal{E}$,
can be approximately separated into a quantum-core-only contribution $E^\mathcal{Q}$,
an analogous contribution of the environment $E^\mathcal{E}$, and an interaction energy $E^{\mathcal{Q}-\mathcal{E}}$,
\begin{equation}
 E_\text{hybrid} = E^\mathcal{Q} + E^\mathcal{E} + E^{\mathcal{Q}-\mathcal{E}} \quad . \label{eq:basic_hybrid}
\end{equation}
When choosing two different methods for the two regions $\mathcal{Q}$ and $\mathcal{E}$, there must not be any electromagnetic interactions included twice or be missing completely. As noted before,
we apply a quantum chemical method for the core region and SFAM for the environment.
Furthermore, we distinguish two schemes for the interaction energy $E^{\mathcal{Q}-\mathcal{E}}$, (a) one in which the electrostatic interaction
is treated by SFAM and (b) one in which it is described quantum mechanically. The former is known as mechanical embedding~(ME) 
and the latter as electrostatic embedding~(EE).
Although it has been demonstrated that EE provides more accurate results than ME, in particular for small QM regions~\cite{rossbach17}, we implemented both embedding schemes
because EE may not always be available for the QM method of choice. In the case of ME, the QM/SFAM energy expression reads,
\begin{equation} \label{eq:qm_sfam_me}
\begin{split}
 E_\text{QM/SFAM}^\text{ME} = \: & E_\text{SFAM}^{\mathcal{Q}+\mathcal{E}} + E_\text{QM}^{\mathcal{Q}} - \sum_{\substack{(A,B) \\ \in \mathcal{Q}}} E^{AB}_r - \sum_{\substack{(A,B,C) \\ \in \mathcal{Q}}} E^{ABC}_\alpha \\
 & - \sum_{\substack{(A,B,C,D) \\ \in \mathcal{Q}}} E^{ABCD}_\theta - \sum_{\substack{(X,B,C,D) \\ \in \mathcal{Q}}} E^{XBCD}_\varphi \\
 & - \sum_{\substack{A \in \mathcal{Q} \\ B \in \mathcal{Q} \\ A \neq B}} \left( E_\text{estat}^{AB} + E_\text{disp}^{AB} + E_\text{rep}^{AB} \right) - \sum_{\substack{(D,H,A) \\ \in \mathcal{Q}}} E_\text{hb}^{DHA} \quad .
\end{split}
\end{equation}
In Eq.~(\ref{eq:qm_sfam_me}), the energy $E_\text{SFAM}^{\mathcal{Q}+\mathcal{E}}$ refers to the SFAM energy of the full system (i.e., $\mathcal{Q}$ and $\mathcal{E}$ combined)
and $E_\text{QM}^{\mathcal{Q}}$ is the electronic energy obtained in a QM calculation for $\mathcal{Q}$,
\begin{equation}
E_\text{QM}^{\mathcal{Q}} = \langle \widehat{H}^{\mathcal{Q}}_{\text{el,ME}}  \rangle
\end{equation}
with the electronic Hamiltonian in atomic units
\begin{equation}
\begin{split}
\widehat{H}^{\mathcal{Q}}_{\text{el,ME}} = 
& - \sum_{i\,=\,1}^{N^\mathcal{Q}_{\text{el}}} \frac{1}{2m_e} \Delta_i
- \sum_{\alpha\,=\,1}^{N^\mathcal{Q}_{\text{nuc}}} \frac{1}{2M_\alpha} \Delta_\alpha
- \sum_{i\,=\,1}^{N^\mathcal{Q}_{\text{el}}} \: \sum_{\alpha\,=\,1}^{N^\mathcal{Q}_{\text{nuc}}} \frac{Z_\alpha}{\vert \mathbf{r}_i - \mathbf{R}_\alpha \vert} \\
& + \sum_{i\,=\,1}^{N^\mathcal{Q}_{\text{el}}} \: \sum_{j\,=\,i+1}^{N^\mathcal{Q}_{\text{el}}} \frac{1}{\vert \mathbf{r}_i - \mathbf{r}_j \vert}
+ \sum_{\alpha\,=\,1}^{N^\mathcal{Q}_{\text{nuc}}} \: \sum_{\beta\,=\,\alpha+1}^{N^\mathcal{Q}_{\text{nuc}}} \frac{Z_\alpha\,Z_\beta}{\vert \mathbf{R}_\alpha - \mathbf{R}_\beta \vert} ,
\end{split}
\end{equation}
where $N^\mathcal{Q}_{\text{el}}$ is the number of electrons, 
$N^\mathcal{Q}_{\text{nuc}}$ is the number of atomic nuclei, 
$Z_\alpha$ is the nuclear charge of nucleus $\alpha$,
and $\mathbf{r}_i$ and $\mathbf{R}_\alpha$ are the Cartesian coordinates 
of electrons and nuclei in $\mathcal{Q}$, respectively.

The latter part of Eq.\ (\ref{eq:qm_sfam_me}) subtracts all energy contributions in the MM force field
that are covered by $E_\text{QM}^{\mathcal{Q}}$, i.e., all bond terms
$E^{AB}_r$ with the atoms $A$ and $B$ both in the QM region, all angle terms with at least two, dihedral terms with at least three,
and all improper dihedral terms with all four of their corresponding atoms in the QM region.
All pairwise noncovalent interaction terms are subtracted for each pair of atoms in $\mathcal{Q}$. Hence, all noncovalent interactions between $\mathcal{Q}$ and $\mathcal{E}$ are described by SFAM, which is a consistent approach if the electronic structure model
does not acount for dispersive interactions (as in many standard density functionals) so that they can be treated semi-classically
everywhere in the system.
For EE, we apply a quantum mechanical description of the electrostatic interaction by defining $E_\text{QM/SFAM}^\text{EE}$ as,
\begin{equation} \label{eq:qm_sfam_ee}
 E_\text{QM/SFAM}^\text{EE} = E_\text{QM/SFAM}^\text{ME} + E^\mathcal{Q}_{\text{QM,\,estat}} - \sum_{\substack{A \in \mathcal{Q} \\ B \in \mathcal{E}}} E_\text{estat}^{AB} \quad ,
\end{equation}
with
\begin{equation}
 E^\mathcal{Q}_{\text{QM,\,estat}} = E_\text{QM,EE}^{\mathcal{Q}} - E_\text{QM,ME}^{\mathcal{Q}} \quad .
\end{equation}
Eq.~(\ref{eq:qm_sfam_ee}) includes an additional energy contribution obtained in the QM calculation of $\mathcal{Q}$, namely $E^\mathcal{Q}_{\text{QM,\,estat}}$.
This is the interaction energy of the elementary particles (electrons and nuclei) in $\mathcal{Q}$ with the electrostatic potential generated by
SFAM's atomic partial charges located at atomic positions in $\mathcal{E}$ as point charges. 
Naturally, the classical equivalent of this interaction must be subtracted to avoid double counting.
The electronic Hamiltonian operator (in Hartree atomic units) is therefore different for EE compared to ME,
\begin{equation}
 \widehat{H}^{\mathcal{Q}}_{\text{el,EE}} = \widehat{H}^{\mathcal{Q}}_{\text{el,ME}} - \sum_{i\,=\,1}^{N^\mathcal{Q}_{\text{el}}} \sum_{A \in \mathcal{E}} \frac{q_A}{\vert \mathbf{r}_i - \mathbf{r}_A \vert}
 + \sum_{\alpha\,=\,1}^{N^\mathcal{Q}_{\text{nuc}}} \sum_{A \in \mathcal{E}} \frac{Z_\alpha\,q_A}{\vert \mathbf{R}_\alpha - \mathbf{r}_A \vert} \quad .
\end{equation}
Here, 
$q_A$ is the partial charge of atom $A$ in $\mathcal{E}$,
and as before, $Z_\alpha$ is the nuclear charge of nucleus $\alpha$ in $\mathcal{Q}$, and 
$\mathbf{r}_A$, $\mathbf{r}_i$, and $\mathbf{R}_\alpha$ are the Cartesian coordinates of the atoms in $\mathcal{E}$ and
the electrons and nuclei in $\mathcal{Q}$, respectively.
The van der Waals interactions are treated at the SFAM level (based on semi-classical dispersion corrections~\cite{johnson09, grimme10, grimme11, grimme11_review} of Grimme) in both embedding schemes.
Within $\mathcal{Q}$, the QM method must take care of dispersive interactions.

\subsection{QM--SFAM boundary}
\label{sec:boundary_treatment}

The challenge of describing a single molecular system with two different physical theories becomes most apparent at the boundary 
of the two regions $\mathcal{Q}$ and $\mathcal{E}$,
particularly if the boundary intersects a covalent chemical bond.~\cite{field90}
Various strategies have been developed for modeling this QM--MM boundary.~\cite{senn06, groenhof13}
The by far most common one is the link-atom approach~\cite{field90, singh86, maseras95, eichler97, antes98, das02, swart03},
in which the covalent bond at the border of the QM region is valence saturated by a hydrogen atom or some other prototypical residue (e.g., a methyl group). 
The most prominent alternative is to generate localized bond orbitals from a slightly larger QM calculation and include these doubly occupied orbitals in the QM calculation of the
hybrid method. During the self-consistent field (SCF) optimization of the orbitals, 
these artificial orbitals are kept frozen.
This approach known as the local-SCF method~\cite{thery94, monard96, assfeld96, ferre02} was introduced by
Rivail and coworkers and later extended by Gao and coworkers~\cite{gao98, amara00, garcia04}. 
Furthermore, advanced embedding approaches may be applied to separate the QM region from an environment,
such as projector-based embedding and embedded mean-field theory~\cite{huzinaga71, Manby2012, fornace15, hegely16, lee19}, frozen density embedding~\cite{wesolowski93, Neugebauer2005, Neugebauer2005a, Neugebauer2005b, Neugebauer2006, Jacob2006, wesolowski08, jacob08, pernal09, fux10, Jacob2014, wesolowski15},
or the subsystem separation by unitary block-diagonalization approach~(SSUB)~\cite{muhlbach18}.

In our QM/SFAM implementation, we focus on the link-atom approach, but emphasize that our implementation can be extended
to include the more advanced boundary schemes mentioned above.
It is crucial for the link-atom approach to carefully select the bonds which indicate the boundary of the QM region, because replacing heavy atoms with hydrogen link atoms introduces artificial effects
on neighboring molecular entities by distorting the electronic structure. 

Our automated strategy for placing link atoms is described in section~\ref{sec:automated_qm_region_selection}.
Along the vector of a single bond $i$, 
\begin{equation}
 \mathbf{r}_{\mathcal{E}_i\mathcal{Q}_i} = \mathbf{r}_{\mathcal{E}_i} - \mathbf{r}_{\mathcal{Q}_i} \quad , \label{eq:bond_vector}
\end{equation}
which is defined by the coordinates of atom $\mathcal{E}_i$ in the environment and 
$\mathcal{Q}_i$ of the quantum region bonded to $\mathcal{E}_i$,
a hydrogen link atom is positioned at 
\begin{equation}
 \mathbf{r}_{\text{H}_i} = \mathbf{r}_{\mathcal{Q}_i} + \left( R_{\mathcal{Q}_i,\text{cov}} + R_\text{H,cov} \right) \frac{\mathbf{r}_{\mathcal{E}_i\mathcal{Q}_i}}{\vert \mathbf{r}_{\mathcal{E}_i\mathcal{Q}_i} \vert}  \quad , \label{eq:link_atom_placement}
\end{equation}
with the covalent radii~\cite{huheey06, greenwood12} $R_{\mathcal{Q}_i,\text{cov}}$ and $R_{H,\text{cov}}$ of $\mathcal{Q}_i$ and hydrogen.
This approach
allows us to cut through single bonds only~(see section~\ref{sec:automated_qm_region_selection}), which is, however, not a
severe restriction, especially considering the fact that the QM region can be enlarged to eventually meet it.

To exploit the force on an artificial link atom H$_i$, which is located between $\mathcal{E}_i$ and $\mathcal{Q}_i$, 
its energy gradient $\mathbf{g}_\text{H$_i$}$
calculated in the QM calculation must be distributed to $\mathcal{E}_i$ and
$\mathcal{Q}_i$. The gradient contributions $\tilde{\mathbf{g}}_{\mathcal{E}_i}$ and $\tilde{\mathbf{g}}_{\mathcal{Q}_i}$ in direction $\mu$ ($\mu=x,y,z$) are, following Ref.~\citenum{walker08},
\begin{align}
  \tilde{{g}}_{\mathcal{Q}_i, \mu} &=
  \mathbf{g}_\text{H$_i$}^\mathrm{T} \cdot \left( \left( 1 - \frac{\vert \mathbf{r}_{\mathcal{Q}_i} - \mathbf{r}_\text{H$_i$} \vert}{\vert \mathbf{r}_{\mathcal{E}_i\mathcal{Q}_i} \vert} \right) \mathbf{u}_\mu + \frac{\vert \mathbf{r}_{\mathcal{Q}_i} - \mathbf{r}_\text{H$_i$} \vert d_\mu}{\vert \mathbf{r}_{\mathcal{E}_i\mathcal{Q}_i} \vert^3} \, \mathbf{r}_{\mathcal{E}_i\mathcal{Q}_i} \right) \quad , \label{eq:grad_shift_1} \\ 
  \tilde{{g}}_{\mathcal{E}_i, \mu} &= \mathbf{g}_\text{H$_i$}^\mathrm{T}\cdot \left( \frac{\vert \mathbf{r}_{\mathcal{Q}_i} - \mathbf{r}_\text{H$_i$} \vert}{\vert \mathbf{r}_{\mathcal{E}_i\mathcal{Q}_i} \vert} \mathbf{u}_\mu - \frac{\vert \mathbf{r}_{\mathcal{Q}_i} - \mathbf{r}_\text{H$_i$} \vert d_\mu}{\vert \mathbf{r}_{\mathcal{E}_i\mathcal{Q}_i} \vert^3} \, \mathbf{r}_{\mathcal{E}_i\mathcal{Q}_i}   \right) \quad . \label{eq:grad_shift_2}
\end{align}
where $d_\mu$ is the absolute value of the difference in the $\mu$-th component of $\mathbf{r}_{\mathcal{E}_i}$ and $\mathbf{r}_{\mathcal{Q}_i}$,
and $\mathbf{u}_\mu$ is a unit vector in direction $\mu$, e.g., $\mathbf{u}_z = \left( 0, 0, 1 \right)^\text{T}$.
With these equations, it is straightforward to calculate analytic gradients for the QM/SFAM energy 
as long as analytic gradients are available for the QM method (including gradients on the external point charges).
These gradients are essential for efficient structure optimizations and for molecular dynamics simulations with our QM/SFAM model.

\begin{figure}[H]
\begin{center}
\includegraphics[width=0.7\textwidth, trim={0 0 0 0},clip]{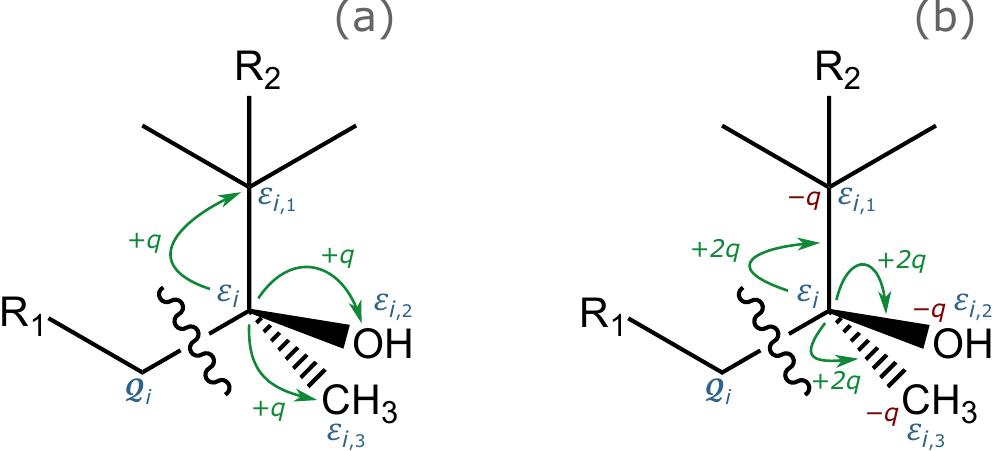}
\end{center}
\vspace*{-0.25cm}
\caption{\label{fig:charge_redistr_schemes}\small Illustration of (a) the redistribution of charge (RC) scheme and (b) the redistribution of charge and dipoles (RCD) scheme to prevent overpolarization of a
link atom (not shown) by the charge on $\mathcal{E}_i$ at the QM/SFAM boundary $i$ (wiggly line). Green color highlights 
the redistribution of partial charges toward the positions to which the arrows are pointing, whereas the color red indicates
a subtraction of charge. We define $q = q_{\mathcal{E}_i}/n$ with $n$ being the number of non-QM neighbors of $\mathcal{E}_i$. 
In this example, $n=3$ 
as the neighboring atoms are
$\mathcal{E}_{i, 1}$, $\mathcal{E}_{i, 2}$, and $\mathcal{E}_{i, 3}$. $R_1$ represents the remainder of the
QM region and $R_2$ denotes the rest of the environment.}
\end{figure}

In electrostatic embedding, another issue arises at the $\mathcal{Q}$--$\mathcal{E}$ boundary. Since the partial charge $q_{\mathcal{E}_i}$
of some atom $\mathcal{E}_i$ at the boundary is included into the QM calculation as an external point charge,
the link atom may suffer from overpolarization effects caused by the close proximity of that charge. 
To counteract this artificial effect, many strategies have been proposed such as deleting the charge~\cite{singh86, eurenius96, ryde96}, redistributing it~\cite{sherwood97, devries99, sherwood03, konig05, lin05},
or smearing it out~\cite{eichinger99, das02, amara03} by replacing the point charge by a Gaussian charge distribution centered at $\mathcal{E}_i$.
We apply a charge redistribution scheme that was shown to produce accurate results compared to full QM calculations~\cite{lin05}.
We implemented two variants of the charge redistribution: 
(i) one in which the total charge is conserved (redistribution of charge, denoted RC) and 
(ii) one in which also the bond dipoles of the first shell of bonds in $\mathcal{E}$ are conserved 
(redistribution of charge and dipoles, denoted RCD).
Both schemes are illustrated in Fig.~\ref{fig:charge_redistr_schemes}.

In both schemes, the charge on $\mathcal{E}_i$ vanishes; i.e., the new charge is $\tilde{q}_{\mathcal{E}_i}  = 0$.
In the RC scheme, the charge is shifted equally to the $n$ neighbors $\mathcal{E}_{i, k}$ (with $k \in \{ 1,\dots, n \}$) 
of $\mathcal{E}_i$, which are also in $\mathcal{E}$,
\begin{equation}
 \tilde{q}_{\mathcal{E}_{i, k}} = \frac{q_{\mathcal{E}_i}}{n} \qquad \text{for} \:\: k = 1,\dots, n \quad .
\end{equation}
In the RCD scheme, to conserve the bond dipoles of the bonds $\mathcal{E}_i - \mathcal{E}_{i, k}$,
the charge on $\mathcal{E}_i$ is shifted to the positions half-way in between the $\mathcal{E}_i - \mathcal{E}_{i, k}$ bond vectors
and doubled in magnitude, resulting in auxiliary charges at positions $\mathbf{r}_{\text{aux},k}$,
\begin{equation}
\tilde{q}_{\mathbf{r}_{\text{aux},k}} = \frac{2\,q_{\mathcal{E}_i}}{n} \quad ,
\end{equation}
with
\begin{equation}
	    \mathbf{r}_{\text{aux},k} = \frac{1}{2} \left( \mathbf{r}_{\mathcal{E}_i} + \mathbf{r}_{\mathcal{E}_{i, k}} \right) \quad .
\end{equation}
The factor of two, i.e., the doubling of the shifted charge, is introduced to preserve the magnitude of the bond dipole as the distance between the charges is halved.
Consequently, also the charges on the neighboring atoms must be adjusted so that the new charges on atoms $\mathcal{E}_{i, k}$ become
\begin{equation}
 \tilde{q}_{\mathcal{E}_{i, k}} = q_{\mathcal{E}_{i, k}} - \frac{q_{\mathcal{E}_i}}{n} \quad .
\end{equation}

\subsection{QM/SFAM structure optimization}

As targets of QM/SFAM are large systems
with many degrees of freedom, structure optimizations tend to require many iterations to reach convergence. 
Most of the degrees of freedom to optimize can be attributed to the environment $\mathcal{E}$ and therefore the necessity to perform a QM calculation
in every optimization step can be avoided by a microiteration-based structure optimization. Several variants of such an algorithm exist and have been implemented in QM/MM programs
to accelerate structure optimizations of large systems~\cite{hall00, vreven03, vreven06, kastner07, melaccio11, metz14, caprasecca14}.
The aim of these approaches is to reach the same local energy minimum structure as in a regular optimization without expensive QM calculations in every step.

In our variant of the algorithm, the Cartesian coordinates of all atoms in $\mathcal{Q}$ as well as those of atoms in $\mathcal{E}$ within a distance $R$ to any atom in $\mathcal{Q}$ are frozen, while all remaining MM degrees of freedom are
	  relaxed (either until convergence or until a maximum of $N_1$ steps is reached). As all of the atoms in $\mathcal{Q}$ are fixed (i.e., their gradients are treated as zero), no QM calculation is
	  necessary during these microiterations. 
We note that, despite the system-focused parametrization of SFAM that can be tailored to the QM model in QM/SFAM, our
attempts to utilize the MM gradient for the whole system in this step were fruitless. In fact, it is this remaining mismatch of SFAM and
QM forces that requires the environment atoms at the QM boundary to be kept frozen.

Then, the complete system is relaxed according to the full QM/SFAM gradients. 
No nuclear positions are constrained and therefore a QM calculation is
	    needed for each evaluation of the complete gradient. 
Once convergence has been reached the optimization terminates.
If convergence cannot be reached after $N_2$ steps, 
one macroiteration step will be completed and the procedure will iterate again starting with the first MM-only step.


For the parameters of this algorithm, 
we found values of $R = 4\:\text{\AA}$, $N_1 = 1000$ and $N_2 = 15$ to perform well in all examples studied in this work,
but they may be adjusted if needed.
For the individual structure optimizations, the algorithms implemented in the SCINE Utilities library~\cite{scine_utils} are applied.

\subsection{QM/SFAM in molecular explorations}
\label{sec:qm_sfam_exploration}

A crucial issue of molecular mechanics models is that their error in the total energy of the system is expected to scale unfavorably with 
system size, e.g., measured in terms of the number of atoms  $N_\text{at}$. While for small systems, energies obtained with
molecular mechanics have been shown to be accurate, especially with system-focused models~\cite{grimme14, vanduyfhuys15, brunken20}, this is not expected for large systems, which we illustrate
by a simple statistical model.~\cite{bryan60} Consider the covalent terms in the total MM energy expression
which is a sum of approximately $3N_\text{at}$ (mostly) independent terms, each with an uncertainty of $\pm\Delta\varepsilon$. 
We can model the estimated error under the ideal assumption of equal
probabilities to either underestimate or to overestimate the energy of a single potential term by $\Delta\varepsilon$. For large $N_\text{at}$, the corresponding binomial distribution can be approximated by the normal
distribution.~\cite{feller08} As a result, we estimate a total error of at least
\begin{equation}
\label{erroreq}
 \Delta E \approx 0.675 \sqrt{3 p \left( 1 - p \right) N_\text{at}}
\end{equation}
to occur with a probability of 50\% applying these simple assumptions with $p$ being the probability of overestimating an individual potential energy by $\Delta\varepsilon$ instead of underestimating it.
For example, for a system with 1000 atoms it is expected that with a probability of 50\%, a total error of at least $18.5\Delta\varepsilon$ will be observed, which scales with $\sqrt{N_\text{at}}$.
Moreover, the MM model may exhibit a systematic over- or underestimation of the potential energies to some (minor) extent, resulting in an expected error that scales linearly with system size.
Furthermore, we note that the MM noncovalent pair interaction terms, which outnumber the covalent terms, are more difficult to
assess with respect to their error contribution\cite{weymuth18,proppe19} because of their distance and hence structure dependence. 

Regardless of the simplified assumptions inherent to Eq.~(\ref{erroreq})
such as neglecting additional uncertainties introduced by the noncovalent interactions (see also Refs. \cite{weymuth18,proppe19}),
it is apparent that for large systems the energies of classical models may exhibit large uncertainties (even with system-focused approaches).
By contrast, atomic forces are local quantities evaluated as partial first-order derivatives at a given reference structure for each atomic nucleus.

In view of these considerations, common practice in QM/MM studies is to generate and sample structures,
either by molecular dynamics simulations or structure optimizations.~\cite{claeyssens06, walker07, marti07, mata08, liao12, polyak12, berraud16}
To obtain accurate energies of local minima on the PES, it is common to freeze all MM atoms beyond a given distance
from the active site during structure optimizations~\cite{bathelt05, mata08, geronimo14, cooper14, finkelmann14} to obtain a converged
structure at smaller computational cost and with larger resemblance of a reference structure such as a structure measured by X-ray diffraction~\cite{ke12},
and to eliminate the contribution of most of the MM region to the total energy.
The advantage of this strategy, compared to neglecting all MM contributions to the total energy, is
that effects of structural changes close to the active site are captured.
However, there exist no standardized guidelines for the choice of this additional cutoff parameter, which may have a significant 
effect on calculated energies.

Considering all of the aforementioned factors, we introduce a reduced QM/SFAM energy $E_\text{QM/SFAM}^\text{red}$ to counteract
the possibly large uncertainties induced by the classical description of a large environment $\mathcal{E}$,
\begin{equation}
 E_\text{QM/SFAM}^\text{red} = E^\mathcal{Q}_\text{QM} + E^{\mathcal{Q}-\mathcal{E}} \quad , \label{eq:reduced_qm_sfam_energy}
\end{equation}
in which any covalent SFAM contributions as well as the noncovalent interactions within the environment
are neglected.
The QM calculation is embedded into the environment by including the $\mathcal{Q}$--$\mathcal{E}$
interaction $ E^{\mathcal{Q}-\mathcal{E}}$
either through mechanical or electrostatic embedding (see section~\ref{sec:qm_mm_formalism}).

We propose that during a molecular exploration, relevant structures should be identified with the complete QM/SFAM model (e.g., by molecular dynamics simulations or structure optimizations),
while for the energy differences of intermediate structures on the PES,
the difference between the two strategies for computing the energy,
\begin{equation}
 \Delta = \Delta E_\text{QM/SFAM} - \Delta E_\text{QM/SFAM}^\text{red} \quad ,
\end{equation}
should be monitored closely.
For $\Delta E_\text{QM/SFAM}^\text{red}$, it is crucial that
energy contributions from structural changes in close proximity to the active site are picked up by the QM calculation.

\subsection{Automated re-parametrization for flexible QM region definitions}
\label{sec:moving_qm_region}

With SFAM as the classical part of the hybrid approach, parameters are always generated for the whole system automatically before starting a molecular exploration.
Therefore, one is not restricted in the selection of the QM region and may freely re-define the QM region. It can be valuable to have this flexibility in 
automated reaction network explorations~\cite{simm19, unsleber20} as well as in reactive molecular dynamics simulations
because reactive centers can shift during a multi-step mechanism.
This feature is also a requirement for applying QM/SFAM in an interactive quantum chemistry framework~\cite{haag14, vaucher16},
as the ability of the operator to choose a region of interest in a large system
should not be limited by missing parameters.

Naturally, QM/SFAM does not require parameters for the covalent terms in the QM region. This means that a reaction that takes place in the QM region and modifies
the local connectivity of the atoms (and hence the SFAM topology), does not result in the model to become unapplicable. Note that parameters for van der Waals interactions,
namely the dispersion coefficients, may be required in the QM region. These can be either quickly re-evaluated or, as an approximation, kept constant even after the modification of the topology,
because the dispersion coefficients for the same types of elements are expected to be similar (we note 
that the dispersion coefficients of the predecessor of D3, i.e., D2,\cite{grimme06}
are fixed for each pair of elements).
Partial charges are expected to be less transferable after a chemical reaction; however, these are not needed for atoms in the QM region.

Even if no bond breaking and bond formation processes are possible
in the classical region, a re-definition of the QM region can move atoms affected by such processes from the QM region to the environment.
In this case, the connectivity of the atoms is modified in the classical region and therefore SFAM must be re-parametrized if the newly required SFAM parameters
are not available due to the existence of the same bonding pattern somewhere else in the system.
To cope with such events, we here extend our SFAM parametrization procedure~\cite{brunken20} by the option to re-parametrize 
locally, for which QM data from the QM-region calculation may be exploited. 
In general, the missing parameters must be obtained in an efficient way at a small fraction of the cost of the full-system parametrization.

\begin{figure}[H]
\begin{center}
\includegraphics[width=0.6\textwidth, trim={0 0 0 0},clip]{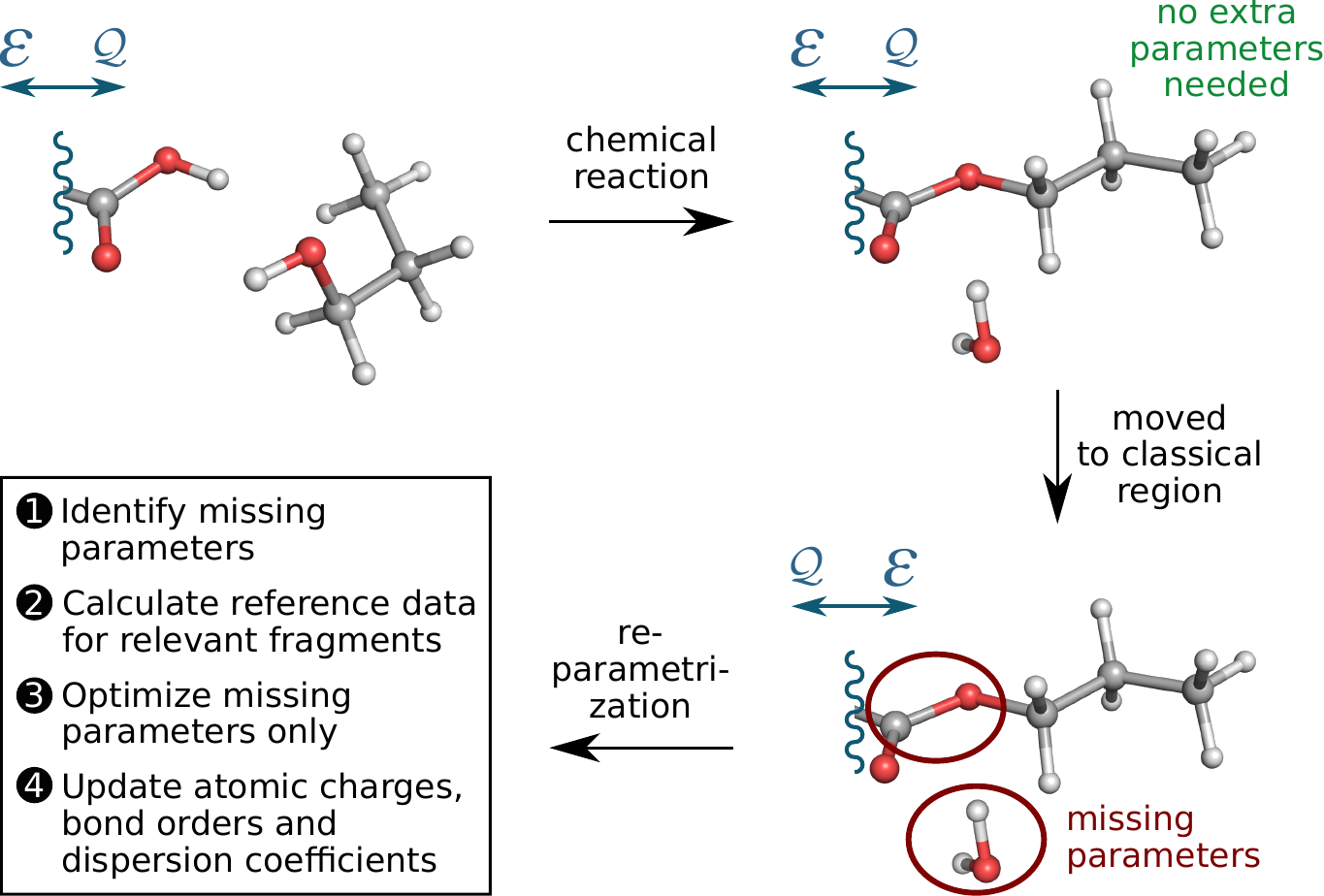}
\end{center}
\vspace*{-0.25cm}
\caption{\label{fig:reparametrization_illustration}\small Local re-parametrization of the SFAM model if a region in which
new covalent bonds are formed or broken is moved out of the QM region $\mathcal{Q}$ to 
become part of the environment $\mathcal{E}$. Shown is an esterification reaction, which is embedded in a
large environment (not shown for the sake of clarity). The box provides a short summary of the steps involved in a local model reparametrization.}
\end{figure}

At the start, we identify all parameters which are not covered by the existing set of SFAM parameters.
To calculate the required reference data~(i.e., optimized local geometries, Hessian matrix, atomic partial charges, bond orders),
we fragment the whole system as explained in our original work on SFAM~\cite{brunken20},
but perform calculations only on those fragments that were generated around the atoms involved in the bonds, angles, or dihedral angles with missing parameters.
Subsequently, the parameters are optimized based on the calculated Hessians and local equilibrium geometries. 
Partial charges and connectivity information
(obtained from Mayer~\cite{mayer83, mayer86} covalent bond orders)
are updated for all atoms and bonds for which new information is available (see our original work on SFAM~\cite{brunken20} for details). Moreover, the dispersion coefficients
are re-evaluated for the whole system due to the negligible additional computational effort associated with it;
see Fig.~\ref{fig:reparametrization_illustration} for an overview of the whole procedure.

Finally, we note that this strategy can be combined with a second approach toward flexible QM regions,
i.e., with adaptive QM/MM schemes for molecular dynamics.~\cite{heyden07, bulo09, mones15, zheng16, duster17} These have been developed
in recent years to allow for moving small molecules (e.g., solvent molecules) from the MM to the QM region (and vice versa) during an MD simulation while preserving a smooth description of the total energy.


\subsection{Automated selection of the QM region}
\label{sec:automated_qm_region_selection}

In this section, we introduce our algorithm for the selection of atoms for the QM region. Once a location of the QM region is provided,
either based on structural characteristics that indicate chemical reactivity or by explicit manual determination,
this location allows us to identify an atom around which the QM region is constructed (called ``the center atom'').
Our aim is to define an automated, universal, and data-driven procedure to find an accurate QM/MM model for the description of the reactive center when compared
to a full QM calculation on the system or, if this is computationally not feasible, to the best possible estimate of that.
First, one needs to define a descriptor to measure the accuracy of a given model. As mentioned
in the Introduction, previous work by Karelina and Kulik\cite{karelina17} applied descriptors based on
charge distribution. However, we focus on the forces on atoms in the proximity to the center atom,
because these relate directly to reasonable structures either in structure optimizations or molecular dynamics simulations.

For long-time molecular dynamics simulations (possibly with large
basis sets), one cannot afford as large of a QM region as, for instance, in structure optimizations that require less than 100 single-point gradient calculations.
Moreover, for small QM regions it is also important to have a systematic approach toward a reliable solution.
We therefore emphasize the importance of automation required to carry out a large number of exploratory 
calculations on candidate models of different size which need to be 
automatically set up, carried out, and then analyzed
(including also the construction of the models) 
for the reliable and autonomous QM region determination to be applicable in a routine fashion.

In the following, we first explain how we construct QM regions around a given center atom automatically.
Then, the selection criteria for the QM region are discussed. Finally, we clarify how to obtain reference data for systems where a full QM calculation is not feasible.

The construction of a QM region with a user-defined center atom represents a task analogous to the fragmentation step in our SFAM model generation.~\cite{brunken20}
In the latter case, we construct one molecular fragment around each atom of the system under subsequent valence saturation 
with hydrogen atoms. This is achieved by first
defining a sphere with radius $r_0$ around a selected atom and adding all atoms within it to the fragment. Second, all covalent bonds which were cut by the sphere's edge are identified
and followed outwards recursively until a covalent bond is reached at which the system can be divided and valence saturated by a
residue (currently, hydrogen-atom saturation has been implemented).
Which bonds are considered cleavable is pre-determined, but can be adapted for a given system.
For biochemical systems, $\text{C}_{\text{sp}^3}-\text{X}$ bonds (with $\text{X} = \text{C, N}$)
can be considered a suitable choice because of their abundance in biological macromolecules. 
We emphasize that 
advanced embedding schemes\cite{huzinaga71, Manby2012, fornace15, hegely16, lee19}, frozen density embedding~\cite{wesolowski93, Neugebauer2005, Neugebauer2005a, Neugebauer2005b, Neugebauer2006, Jacob2006, wesolowski08, jacob08, pernal09, fux10, Jacob2014, wesolowski15, fornace15, hegely16, muhlbach18, lee19}
may have the potential to replace this rule-based saturation approach.
Combined with an initial radius between $r_0 = 5.5$\:\AA~and $r_0 = 7.0$\:\AA, our strategy resulted in a maximum fragment size of
under 150~atoms for several of our example systems~\cite{brunken20}, which means that the required reference data can easily be calculated for these fragments with contemporary density functional theory.

To sample several model sizes and boundaries, we introduce a stochastic element to our automated QM/MM model construction. 
If a cleavable bond is reached, the system may be chosen to be split into QM and MM parts at that bond with 
some probability $p$.
Naturally, the resulting set of QM regions will contain duplicates because each QM region is constructed independently.
Hence, the set of QM regions must be deduplicated.
This straightforward approach is chosen over a systematic generation of all possible QM regions, because of its simple implementation in
the current fragmentation framework and the otherwise exploding number of possible QM regions of varying size, for which an exhaustive generation and selection process becomes unfeasible.
Nevertheless, we point out that for small QM regions, the stochastic approach is also capable of generating all possible QM regions exhaustively up to a given size due to the efficiency of the fragmentation
algorithm. Hence, the parameters $r_0$ and $p$ allow for adjusting the QM region size and the variation of sizes. 

To filter and categorize the generated QM regions, we introduce two additional descriptors. The first one is the number of covalent bonds $m_\text{link}$
cut in the process of defining the QM region, which is equal to the number of link atoms in the resulting fragment.
As a second feature of the generated QM regions, we introduce a symmetry
measure $m_\text{sym}$,
\begin{equation}
 m_\text{sym} = \frac{\overline{r}_\text{LDM}}{\overline{r}_\text{MDQ}} \quad , \label{eq:symm_score}
\end{equation}
where $\overline{r}_\text{LDM}$ is the mean distance of the central atom to the three least distant MM atoms~(LDM) and $\overline{r}_\text{MDQ}$
is the mean distance of the central atom to the three most distant QM atoms~(MDQ).
This descriptor can be applied to assess the extent to which atoms are arranged aspherically around the reactive center.

To measure the reliability of some automatically produced QM/MM model $m$ in terms of how accurately the atomic forces 
in close proximity to the center atom are described compared to a reference ('ref'),
we first select a set of $N_\text{repr}$ representative atoms that are closer than a cutoff of $r_\text{repr}$ to the center atom.
The mean absolute error $\varepsilon_k^m$ of the force components ($f_{x,m,k}$, $f_{y,m,k}$ and $f_{z,m,k}$) for a given atom $k$ in this set is given by
\begin{equation}
 \varepsilon_k^m = \frac{1}{3} \Big( \vert f_{x,\,\text{ref},\,k} - f_{x,\,m,\,k} \vert + \vert  f_{y,\,\text{ref},\,k} - f_{y,\,m,\,k} \vert
 + \vert f_{z,\,\text{ref},\,k} - f_{z,\,m,\,k} \vert \Big) \quad . \label{eq:mae_on_force}
\end{equation}
The overall accuracy can then be measured by the mean of these errors,
\begin{equation}
 \varepsilon_\text{mean}^m = \frac{1}{N_\text{repr}} \sum_{k\,=\,1}^{N_\text{repr}} \varepsilon^m_k \quad . \label{eq:mean_error_forces}
\end{equation}

The reference forces $\mathbf{f}_{\text{ref},\,k} = \left(f_{x,\,\text{ref},\,k},\, f_{y,\,\text{ref},\,k},\, f_{z,\,\text{ref},\,k} \right)^\mathrm{T}$ in Eq.~(\ref{eq:mae_on_force})
may be obtained from a QM calculation on a significantly larger system. As only one single-point gradient evaluation is necessary for this purpose, this will be feasible for systems of several
hundred atoms.
Alternatively, we may obtain a reliable estimate for the reference in the case of larger systems by averaging the force vectors obtained from
a sample of $N_\text{ref}$ QM/SFAM models with large QM regions by choosing a radius $r_0$
that is as large as possible for a single-point calculation in order to be still feasible in a reasonable amount of time.
It is important that this estimate is not based on a single reference model, but on many different ones, because it has been shown that one cannot be certain
that molecular properties are converged even with QM regions of up to several hundred atoms.~\cite{sumowski09, flaig12, rossbach17, kulik16}
The comparison of several QM/MM models with different QM/MM boundaries allows to detect whether the atomic forces are converged with the QM region size that was chosen for the reference.
If significant deviations exist between reference calculations, it can be detected and flagged by the algorithm automatically.

Finally, we note that it has been demonstrated that one can deploy the domain-based local pair natural orbital coupled cluster methods~\cite{neese09, neese09_2, hansen11}
as the QM part (allowing for large QM regions) to obtain accurate reaction barrier heights~\cite{claeyssens06, bistoni18},
which is crucial in mechanistic explorations for the subsequent kinetic analysis~\cite{proppe18}.
Running QM/SFAM calculations with these QM methods is enabled through an interface to the quantum chemistry software ORCA~\cite{neese12, neese18}.

\section{Results}
\label{sec:examples_and_results}

We demonstrate our automated QM/SFAM set-up algorithm with two examples.
For the first example, chain A of the peptide hormone insulin~\cite{sonksen00} was chosen because its size of slightly more than 300 atoms
is large enough to test the effects of different QM regions on the QM/SFAM results while at the same time
full QM calculations are still feasible so that a well-defined full-QM reference is available. 
The initial structure was taken from the Protein Data Bank~\cite{bernstein77} (PDB ID: 1AI0).

\subsection{Construction of structural models}
\label{sec:example_setup}

\begin{figure}[H]
\begin{center}
\includegraphics[width=\textwidth, trim={0 0 0 0cm},clip]{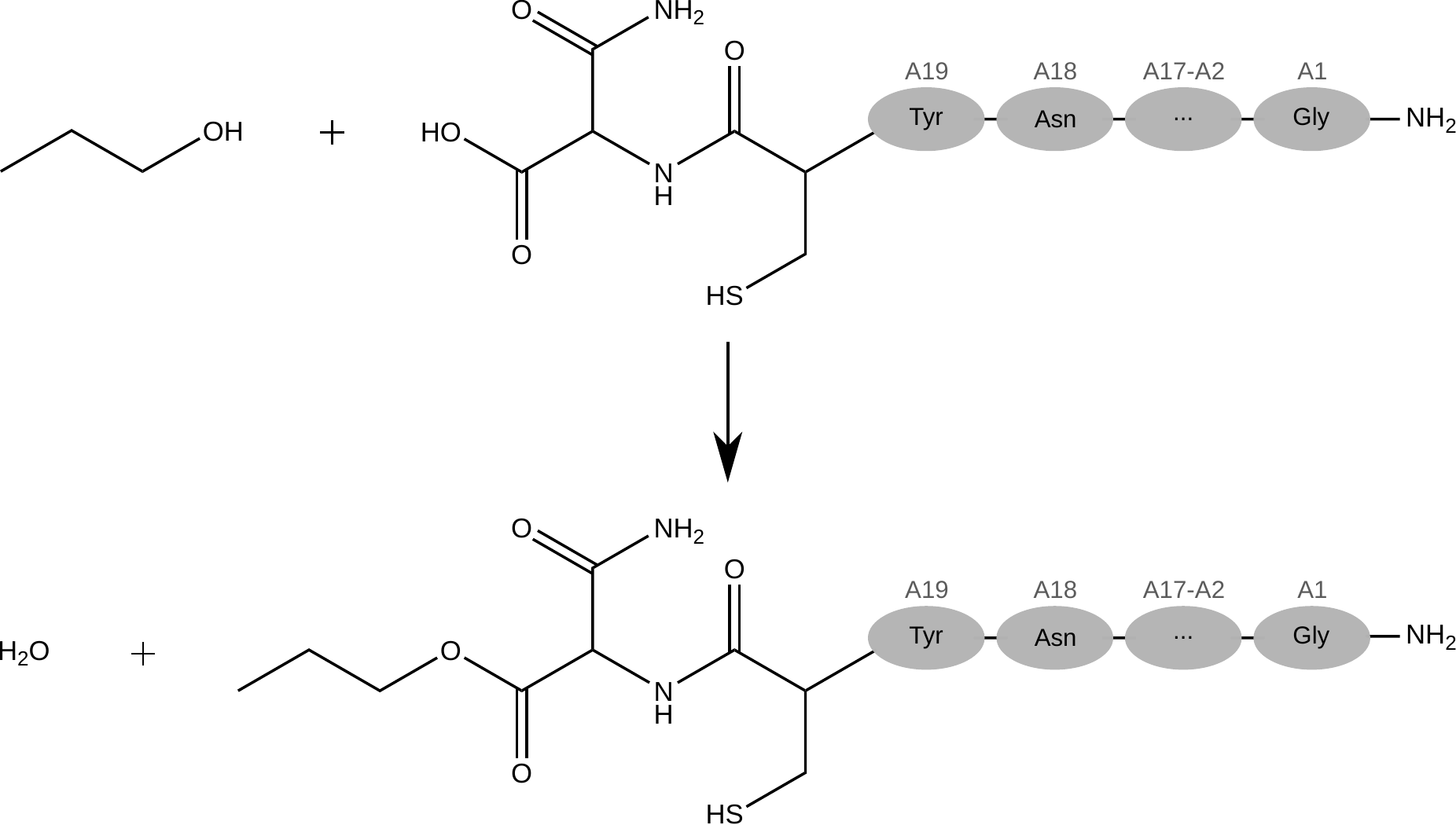}
\end{center}
\vspace*{-0.25cm}
\caption{\label{fig:esterification_scheme}\small Reaction of 1-propanol with chain A of the peptide hormone insulin, which contains 21 amino acids, labeled A1-A21.
The residues A20 (cysteine) and A21 (asparagine) are represented as Lewis structures.
Note that this representation refrains from showing the disulfide bridge connecting two cysteine residues at positions A7 and A11.}
\end{figure}

To study a chemical reaction in this system, we added a 1-propanol molecule in close proximity to the carboxylic acid group of the C-terminus of the chain~(asparagine, A21),
which serves as the initial structure of an esterification reaction. The product structure therefore contains a free water molecule and the propyl ester compound. This reaction is depicted in
Fig.~\ref{fig:esterification_scheme}. With the added alcohol as reactant, the system consists of 328 atoms. We fully pre-optimized the reactant and product structures
with the PM3 semi-empirical method~\cite{stewart89_1, stewart89_2} applying the \texttt{ORCA 4.2} quantum chemistry software.~\cite{neese12, neese18} 
The subsequent DFT optimization with RI-PBE-D3(BJ)/def2-SVP~\cite{whitten73, dunlap79, vahtras93, perdew96, grimme10, weigend05} was limited
to ten optimization steps in order to obtain forces on all atoms that neither vanish nor acquire artificially large numerical values.
The coordinates of these structures can be found in the Supporting Information.

For the reactant structure, a SFAM molecular mechanics model was parametrized in a fully automated fashion~\cite{brunken20}.
The reference data, i.e., optimized structures, Hessians, Mayer bond orders, and 
atomic partial charges, were obtained 
for the RI-PBE-D3(BJ)/def2-SVP electronic structure model with \texttt{ORCA 4.2} driven by our software.
\texttt{ORCA} Hirshfeld charges~\cite{hirshfeld77},
were converted to Charge Model 5 charges by our implementation of the published algorithm~\cite{marenich12} in our SCINE software~\cite{scine}.

\begin{figure}[H]
\begin{center}
\vspace*{-0.5cm}
\includegraphics[width=\textwidth, trim={0 0 0 0cm},clip]{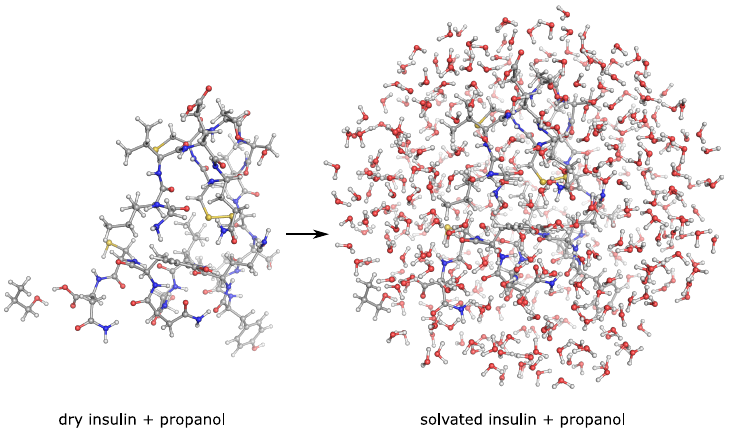}
\end{center}
\vspace*{-0.2cm}
\caption{\label{fig:insulin_molecular_structure}\small 
Dry (left) and microsolvated (right; 16~\AA~water-molecule sphere) insulin with propanol as reactants (cf. Fig.~\ref{fig:esterification_scheme}) 
Atom coloring: carbon in gray, hydrogen in white, oxygen in red, nitrogen in blue, and sulfur in yellow.}
\end{figure}

For the generation of the second example, we solvated the (dry) insulin peptide structure with water molecules. We applied the solvation tool of the \texttt{ADF 2016.107} graphical user interface~\cite{adf}
with which 418 water molecules were added (16~\AA~sphere, solute factor of 2.0), resulting in the second system that now comprises 1582 atoms in total. After the water molecules were added,
the system was not relaxed. Again, we note that this is done deliberately to work with non-zero forces.
The coordinates of the solvated structures can also be found in the Supporting Information (see Fig.~\ref{fig:insulin_molecular_structure}
for a ball-and-stick representation of both systems).
The SFAM parametrization was carried out analogously to the unsolvated insulin system.

We note that for the parametrization, the bottleneck with respect to computing time is the reference data generation step with all other tasks being completed
in less than five minutes for the dry insulin peptide structure and less than one hour for the solvated structure on a modern computing architecture with a single core.
However, the time needed for the reference data generation can vary significantly depending on the number of cores chosen for
parallel execution.
In our set-up, all reference calculations for the solvated structure were completed within a few days in 200 parallel calculations on 4 cores each.

For QM/SFAM calculations, the RI-PBE-D3(BJ)/def2-SVP~\cite{whitten73, dunlap79, vahtras93, perdew96, grimme10, weigend05}
combination of density functional and basis set was applied for the QM region 
through an interface to the \texttt{ORCA 4.2} software~\cite{neese12, neese18} in both examples. 

\subsection{Dry insulin}
\label{sec:example_small}

We first apply the QM-region generation algorithm of
section~\ref{sec:automated_qm_region_selection}
to the structural model of dry insulin.
The carbon atom of the carboxylic acid group of residue A21 (see Fig.~\ref{fig:esterification_scheme}) was chosen as the center atom around which the QM region is constructed.
We study whether the accuracy of the atomic forces close to the center of the QM region can be exploited to ensure 
reproducibility of the QM reference by the QM/SFAM model. 
For this purpose, a large variety of different QM regions was generated, including aspherical ones with a large value of $m_\text{sym}$ (see Eq.~(\ref{eq:symm_score})).
We applied a small initial radius $r_0 = 5.5$~\AA~and a small cutting probability of 15\% until 2000 QM/SFAM structural
models were generated.
The largest QM region obtained contained 328 atoms, which is identical to the whole system (full QM calculation), while the smallest QM region consists of only 32 atoms. 
More information about the distribution of QM region sizes and the distribution of symmetry scores $m_\text{sym}$ obtained by this algorithm can be found in the Supporting Information.

\begin{figure}[H]
\begin{center}
\vspace*{-0.5cm}
\includegraphics[width=0.85\textwidth, trim={0 0 0 0cm},clip]{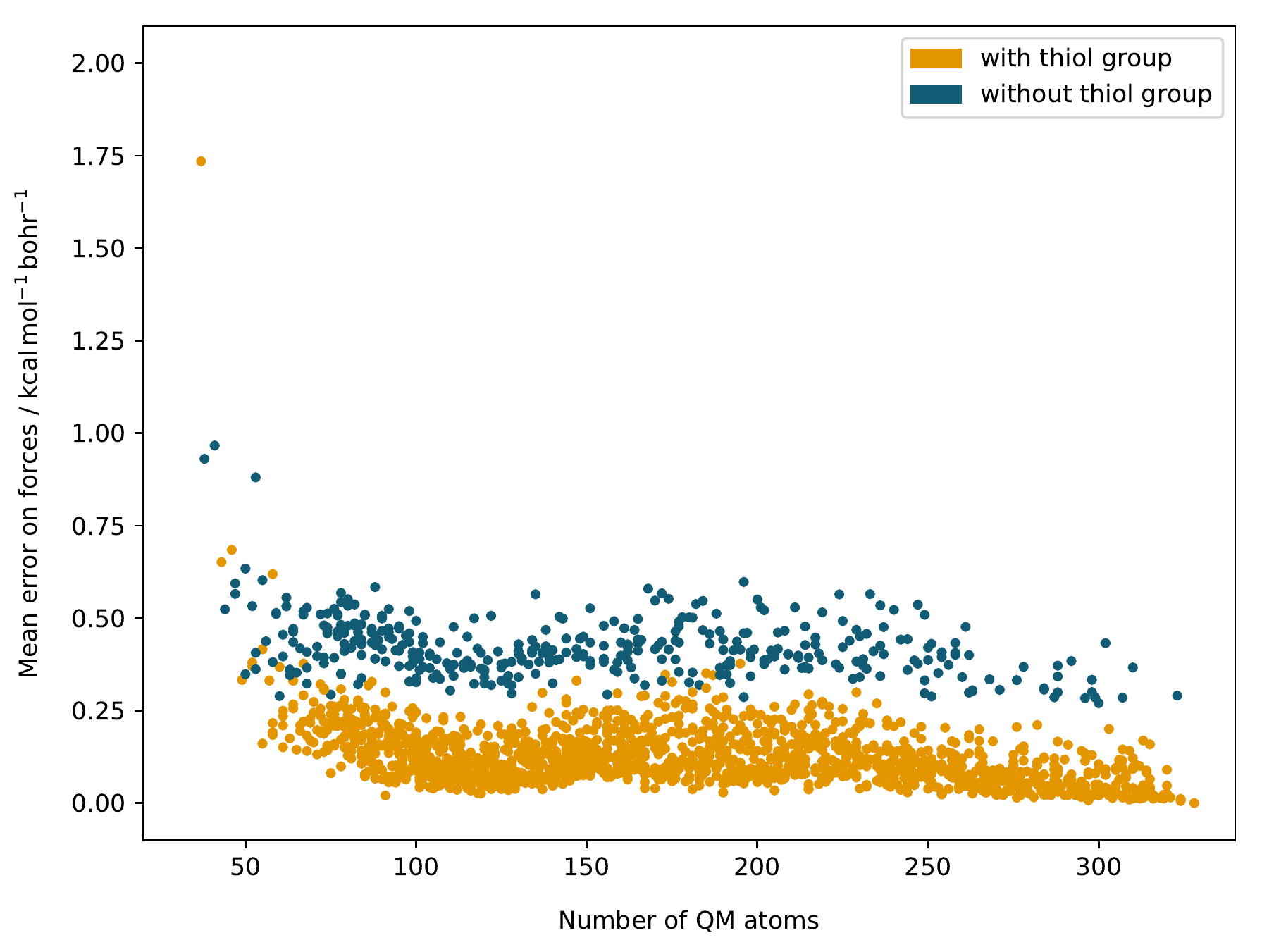}
\end{center}
\vspace*{-0.5cm}
\caption{\label{fig:mean_errors_forces_start}\small Mean error $\varepsilon_\text{mean}^m$ of atomic forces of all non-hydrogen atoms within 4.0\:\AA~of the central carbon atom in the generated
QM/SFAM models for the unsolvated \textit{initial} structure of insulin. Each point corresponds to a different QM/MM model $m$.
The points are colored such that for the orange ones
the thiol group SH$_{5.7}$ is part of the QM region, whereas for the blue points it is part of the environment. 
The point farthest to the right corresponds to the model representing the full QM calculation,
which is taken as the reference for the error evaluation.}
\end{figure}

Fig.~\ref{fig:mean_errors_forces_start} presents the results of evaluating the mean error $\varepsilon_\text{mean}^m$ on the atomic forces of all
non-hydrogen atoms within $r_\text{repr}=4.0$~\AA~of the central carbon atom~($N_\text{repr}=10$), according to Eq.~(\ref{eq:mean_error_forces}). We observe that only four models exhibit a large error of more than 0.75~kcal\,mol$^{-1}$\,bohr$^{-1}$. All of these models consist of less than 60 QM atoms. Despite the obvious rationale that more QM atoms should result in a higher accuracy, the mean error on the forces does not strictly decrease with the size of the QM region. This can be attributed to our choice to generate models with a large variance in 
asphericity (achieved by applying a small value for $p$ and measured by the symmetry score $m_\text{sym}$). 
Large QM regions with more than about 200 QM atoms may be lacking a residue in close proximity to the center of the QM region, which would then result in a large error $\varepsilon_\text{mean}^m$. However, for the models with the smallest error for a given size of the QM region, 
the trend of a decreasing error is observed for models with more than 150 QM atoms. QM regions that are very aspherical may be regarded as unsuitable and should not be considered in an application. This can be achieved by choosing a larger value for $p$ as well as by directly rejecting aspherical models with an improper value for $m_\text{sym}$. We demonstrate this with our second example, the solvated insulin in section \ref{sec:example_large}, for which the expected trend of decreasing $\varepsilon_\text{mean}^m$ with increasing QM region size can be observed (see below).

Moreover, it can be easily seen in Fig.~\ref{fig:mean_errors_forces_start} that the data split into two groups, which are separated by 
about 0.3~kcal\,mol$^{-1}$\,bohr$^{-1}$. It follows from this observation that $\varepsilon_\text{mean}^m$ allows us to easily eliminate one of these groups from our consideration.
The cause for this effect can be attributed to including or excluding the thiol group of the cysteine residue A20 from the QM atoms.
The distance of the sulfur atom of this group to the central atom was 5.7~\AA, which implies that
it was not always included in the QM region. However, it was close enough
to affect the esterification reaction significantly. We call this group SH$_{5.7}$. The coloring in Fig.~\ref{fig:mean_errors_forces_start} highlights this observation. It demonstrates that our descriptor $\varepsilon_\text{mean}^m$ is able to clearly distinguish QM/MM models 
in which SH$_{5.7}$ is part of the QM region from those where it is not. 98.4\% of all models $m$
that contain the SH$_{5.7}$ in their QM region were able to reproduce the reference forces of the full QM calculation with a 
mean error $\varepsilon_\text{mean}^m$ less than 0.3~kcal\,mol$^{-1}$\,bohr$^{-1}$, while all
of the other models produced a larger error than 0.27~kcal\,mol$^{-1}$\,bohr$^{-1}$. With this example, in which the crucial functional group can be easily identified,
we understand that it is possible to automatically and reliably sort out
QM/SFAM models where an unreliable choice of QM region leads to large errors in atomic forces.
Furthermore, we identified a second functional group (a carboxylic acid moiety~(COOH$_{9.5}$) 
at a distance of 9.5~\AA~to the central carbon atom), for which an effect on the forces is observed. As this residue has a larger distance to the reaction center, its influence on the latter is smaller, which results in the observation that the corresponding groups of data are not well separated.
Due to the small size of this effect, we refer to the Supporting Information for its visualization (differently colored version of Fig.~\ref{fig:mean_errors_forces_start}).

\begin{figure}[H]
\begin{center}
\vspace*{-0.5cm}
\includegraphics[width=0.85\textwidth, trim={0 0 0 0cm},clip]{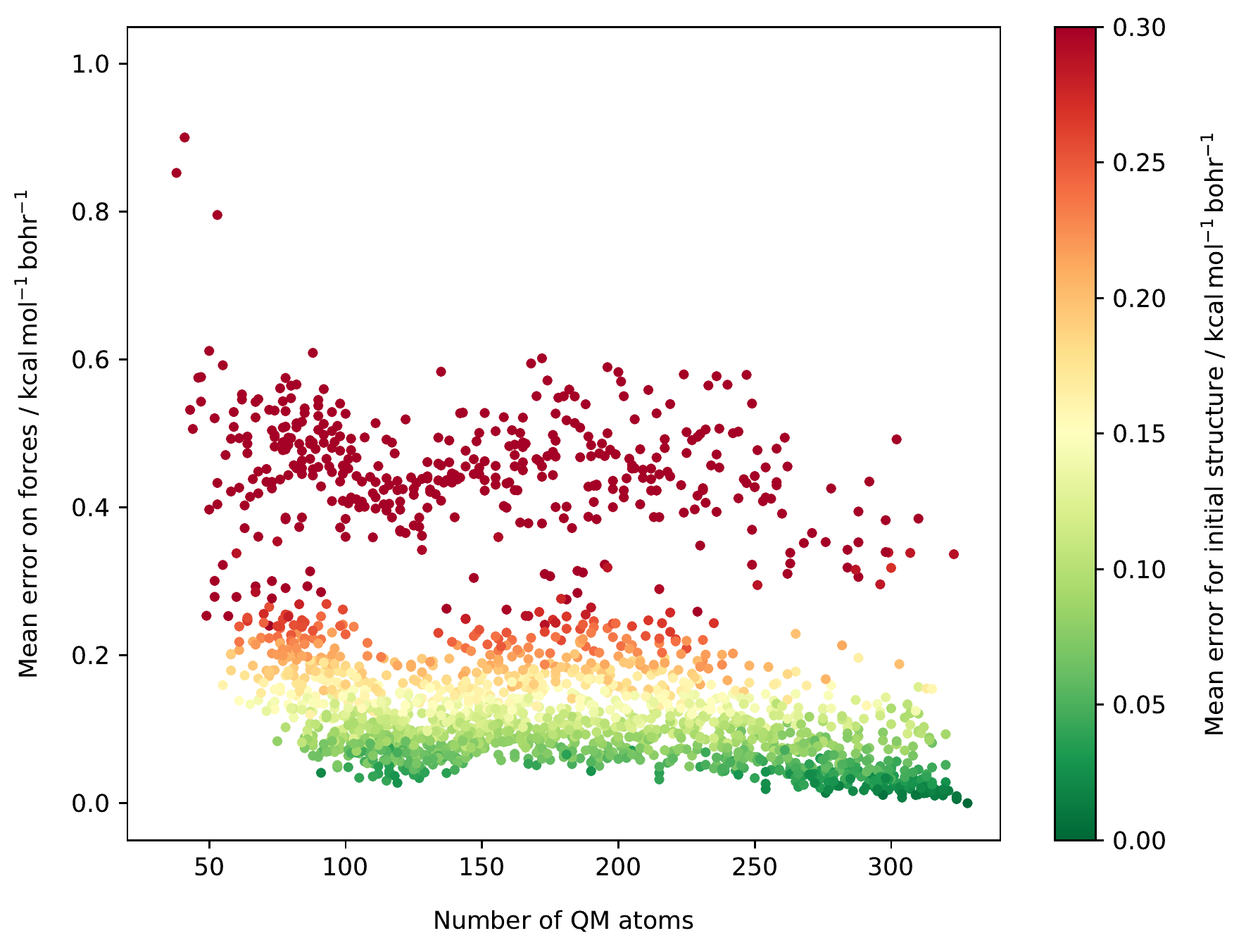}
\end{center}
\vspace*{-0.5cm}
\caption{\label{fig:mean_errors_forces_compare_start_end}\small Mean error $\varepsilon_\text{mean}^m$ on atomic forces of all non-hydrogen atoms within 4.0\:\AA~of the central carbon atom for the generated
QM/SFAM model for the unsolvated insulin \textit{final (product)} structure.
The model with the smallest QM region exhibits a significantly larger error than
1~kcal\,mol$^{-1}$\,bohr$^{-1}$ (as in the case of the initial reactant structure, see Fig.~\ref{fig:mean_errors_forces_start})
and therefore has been omitted here. The colors encode the mean error that was obtained for the \textit{initial} reactant structure
with that model~(i.e., the vertical axis in Fig.~\ref{fig:mean_errors_forces_start}). Models with errors larger than 0.3~kcal\,mol$^{-1}$\,bohr$^{-1}$ on the initial structure
are not further distinguished in color as these are not considered reasonable model candidates.
This representation highlights that the accuracy found for the \textit{initial} structure is matched by
the accuracy obtained at a different location on the PES~(here, the \textit{final} product structure of the esterification reaction).}
\end{figure}

If a QM/SFAM model is able to accurately describe the forces in the reactant structures, but poorly for intermediates, transition states, or products, it will not be sensible to rely on this descriptor for the evaluation of the QM region selection protocol.
Therefore, we evaluated the atomic forces on the same atoms as before for the final (product) structure of the reaction
shown in Fig.~\ref{fig:esterification_scheme} by applying the same 2000 automatically selected QM/MM models
to assess whether the models that led to a small error for the initial structure also performed well for
the final structure. The results of this comparison are presented in Fig.~\ref{fig:mean_errors_forces_compare_start_end},
in which we encode the accuracy of the models on the initial structure forces by their color. We observe an almost perfect
agreement of force deviations measured in terms of $\varepsilon_\text{mean}^m$ for the initial structure and those
for the final structure, indicating that our measure for the reliability
of a selected QM region is likely to be transferable across a PES, at least for close-to-minimum energy structures.

\begin{figure}[H]
\begin{center}
\vspace*{-0.5cm}
\includegraphics[width=0.85\textwidth, trim={0 0 0 0cm},clip]{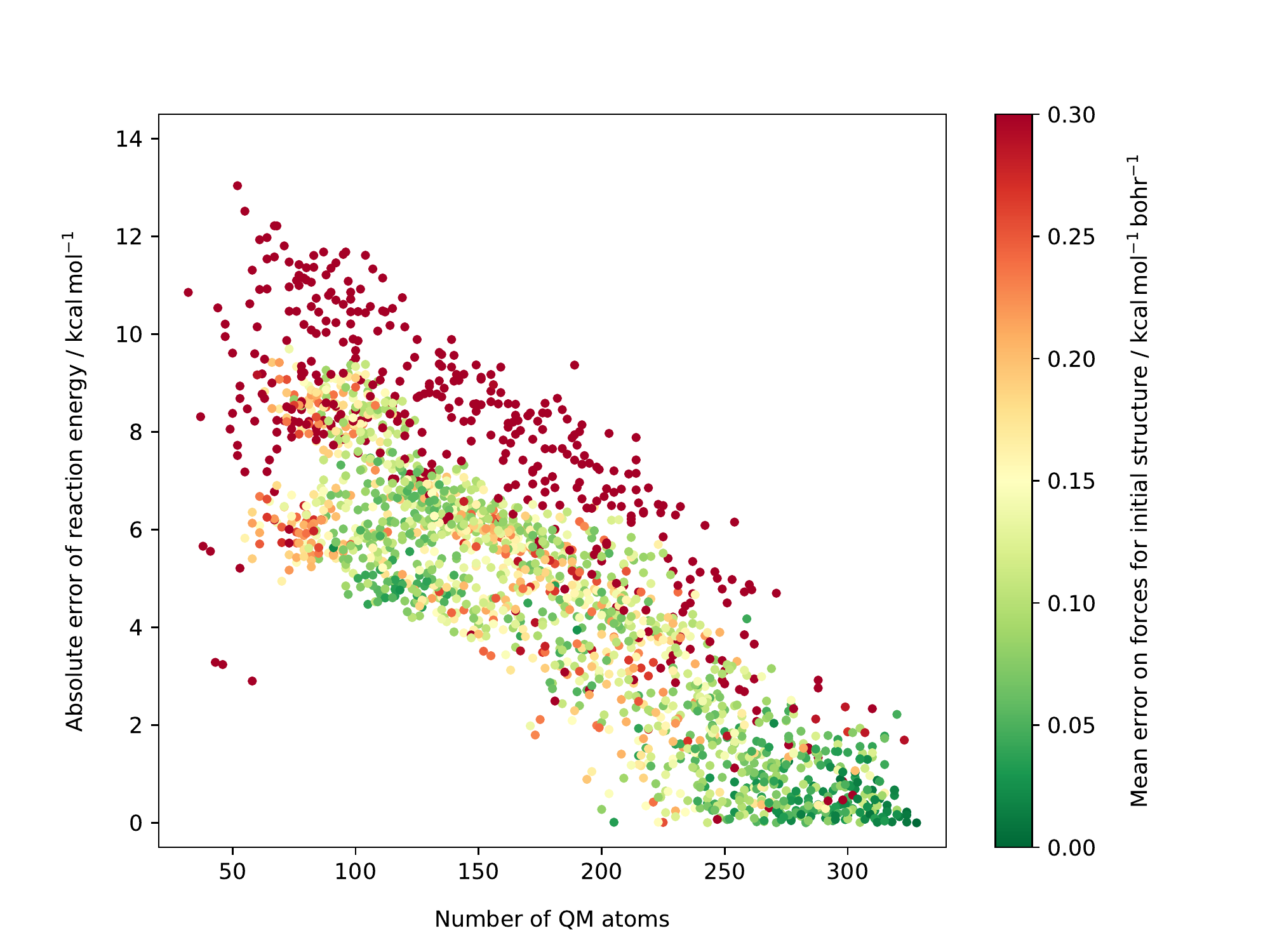}
\end{center}
\vspace*{-0.5cm}
\caption{\label{fig:errors_energy_compare_start}\small Absolute error of the esterification reaction energy for each of the 
generated QM/SFAM models of the unsolvated chain A of insulin.
The colors correspond to the mean error on the atomic forces close to the reaction center obtained for the initial structure with that model~(i.e., the vertical axis in Fig.~\ref{fig:mean_errors_forces_start}).
Models with errors larger than 0.3~kcal\,mol$^{-1}$\,bohr$^{-1}$ on the initial structure
are not further distinguished by color.
}
\end{figure}

Naturally, the size of the QM region may be different for other physical quantities and our assumption that the forces are 
most crucial for making a decision on the size, albeit reasonable from a structural point of view, needs to be scrutinized.
Therefore, we now discuss whether those QM/SFAM models that most accurately reproduced forces also deliver reliable energies. 
For the esterification reaction in dry insulin, the reaction energy calculated as the difference of reduced QM/SFAM energies (see section~\ref{sec:qm_sfam_exploration},
Eq.~(\ref{eq:reduced_qm_sfam_energy})) obtained for the product and reactant structures are presented in Fig.~\ref{fig:errors_energy_compare_start}. where
the same coloring scheme used for the forces of the product structure in Fig.~\ref{fig:mean_errors_forces_compare_start_end}
is applied.

First, we observe that the average energy error is decreasing continuously with growing QM region size.
Second, the models that exhibited a large error on the forces (larger than 0.3~kcal\,mol$^{-1}$\,bohr$^{-1}$)
are separated from the well-performing QM/SFAM models with only a small number of exceptions.
This shows that models that were discarded after evaluating their accuracy on the forces
are also expected to generate large errors in energy, confirming the reliability of our selection strategy.
However, we also observe that for the well-performing models for which the reaction energy only fluctuates by less than 2~kcal\,mol$^{-1}$ for a given QM region size,
the accuracy of the forces does not map perfectly to the accuracy of the reaction energies (compared to Fig.~\ref{fig:mean_errors_forces_compare_start_end}).
Models with differences of less than 0.2~kcal\,mol$^{-1}$\,bohr$^{-1}$ on the forces are not distinguished in terms of the energies.
However, we still reduce the error (compared to the most accurate QM/SFAM model for a given QM region size)
significantly by excluding the models with large errors on the forces; for instance, it is reduced
from roughly 4~kcal\,mol$^{-1}$ to 2~kcal\,mol$^{-1}$ considering QM region sizes between 100 and 200 atoms.

Fig.~\ref{fig:errors_energy_compare_start} shows that the model evaluation based on atomic forces can predict which models exhibit large errors of reaction energy. However, we observe that it cannot be guaranteed that the models with the smallest values of $\varepsilon_\text{mean}^m$ also exhibit the smallest energy errors.
Considering these observations, we conclude that the descriptor $\varepsilon_\text{mean}^m$ can reliably eliminate choices of 
QM regions, which are lacking residues that significantly affect key physical quantities of the reaction. However, the fact that this descriptor is based on a single-point property (the atomic forces), results in both a crucial advantage and a drawback of the method. On the one hand, it allows us to efficiently test many model candidates in an automated fashion against reference data that is calculated with models with very large QM regions. 
On the other hand, we have shown that it cannot be guaranteed that the models with the smallest $\varepsilon_\text{mean}^m$ provide reaction energies that are within 1~kcal\,mol$^{-1}$ of the QM reference. Fig.~\ref{fig:errors_energy_compare_start} demonstrates that in our example, this accuracy can only be achieved by applying very large QM regions with more than 250 atoms. During an exploration of a molecular reaction, we therefore stress the importance of applying models of several QM region sizes in single-point energy calculations to probe for convergence in order to closely monitor the uncertainties by which the QM region selections are affected. Respective algorithmic procedures 
can be included in automated workflows. 

The energy errors discussed so far are, of course, given with respect to a DFT reference and therefore affected with some unknown
uncertainty. We stress, however, that accurate quantum chemical methods, such as coupled cluster approaches \cite{claeyssens06, bistoni18} 
with sufficiently high excitation rank and decent one-particle bases combined with basis-set exploration or explicit correlation factors,
can be applied to obtain more accurate energies. Finally, we point out that it may be beneficial to extend our QM region selection process by explicitly adding energies differences of two or more structures to our descriptor. Hence, we designed our implementation in a modular fashion to allow for such extensions easily.

\subsection{Solvated insulin}
\label{sec:example_large}

At the example of solvated insulin, we demonstrate that it is possible to obtain reliable reference forces even 
for large systems for which one cannot routinely perform a
full QM calculation on the whole system. As described in section~\ref{sec:example_setup}, the structures of the previous example after solvation with water are applied for this purpose.

For this solvated insulin structure, we do not want to generate a variety of different QM regions that is as large in number
as the one in section~\ref{sec:example_small}, because (i) this increases the total number of candidate models
that needs to be tested (which will not be practical in a routine QM/SFAM application) due to the larger total system size,
and (ii) the first example already demonstrated that very aspherical QM regions (e.g., large QM regions without the nearby SH$_{5.7}$ group) do not
provide accurate results. Hence, we set a larger probability of $p = 0.9$ for this example.
The QM/SFAM models were generated by sampling 200 QM regions for a given radius $r_0$ while increasing this value in steps of 0.1~\AA~starting at 5.5~\AA~and terminating at 11~\AA, yielding 11000 models in total.
After the deduplication process, 
673 unique QM/SFAM models were created. The smallest QM region comprised 68 atoms, the largest 410 atoms.
The obtained distribution of QM region sizes is provided in the Supporting Information.

\begin{figure}[H]
\begin{center}
\vspace*{-0.5cm}
\includegraphics[width=0.85\textwidth, trim={0 0 0 0cm},clip]{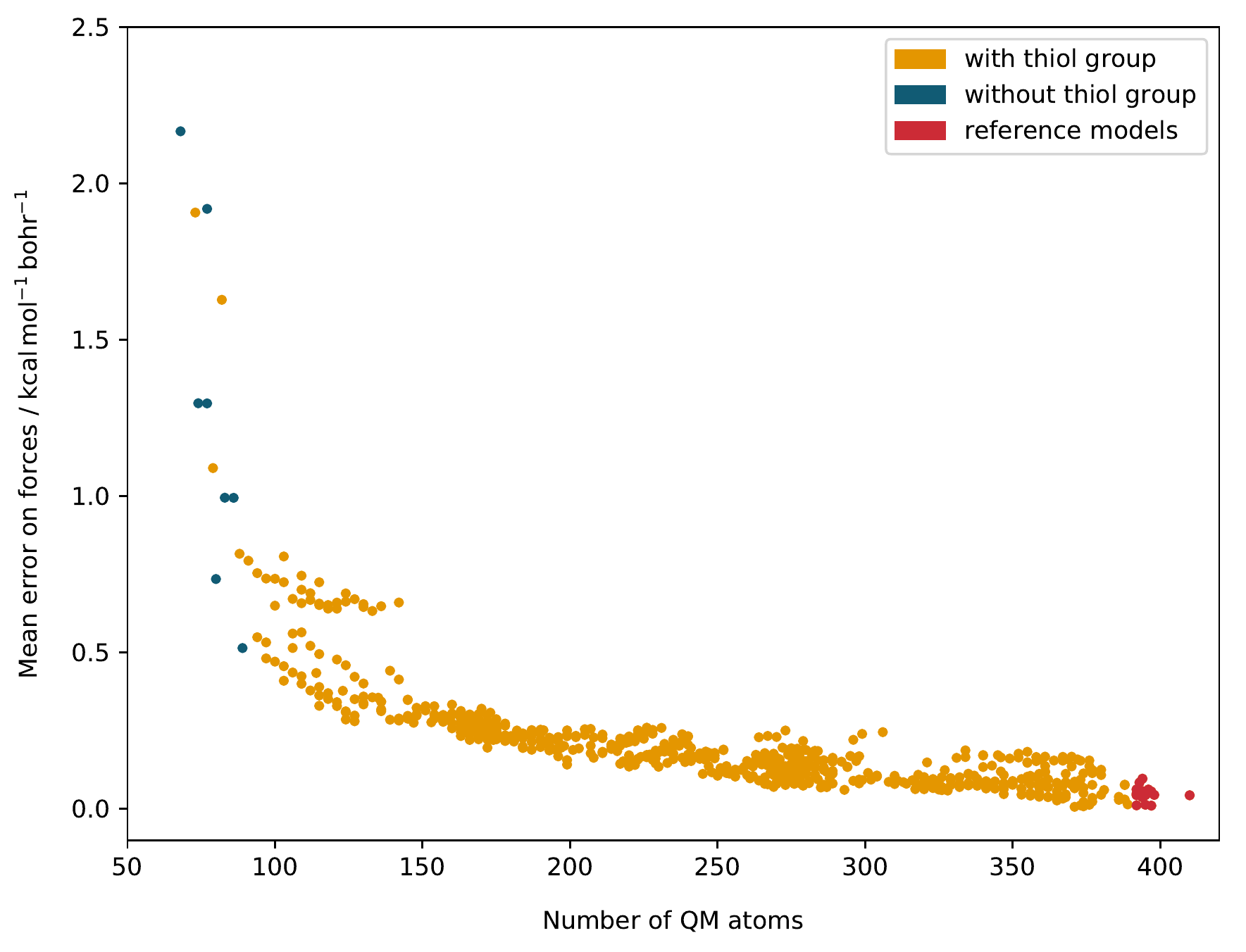}
\end{center}
\vspace*{-0.5cm}
\caption{\label{fig:mean_errors_forces_solvated_start}\small Mean error 
$\varepsilon_\text{mean}^m$
on the atomic forces of all non-hydrogen atoms within 4.0\:\AA~of the central carbon atom for the generated
QM/SFAM models for the initial structure of solvated insulin. Each point corresponds to a different QM/SFAM model.
The points are colored such that for the orange ones
the thiol group SH$_{5.7}$ is part of the QM region, while for the blue points it is part of the environment. The red points correspond to the models for which the mean of the forces was taken as the reference
(QM regions of at least 390 atoms). With the fragmentation settings applied here, which differ from those of 
unsolvated insulin, all models with more than 100 QM atoms include the thiol group into the QM region.}
\end{figure}

As described in section~\ref{sec:automated_qm_region_selection}, we take the mean of $N_\text{ref}$ models with very large QM regions to obtain the reference forces.
In this case, we assign all models with QM regions of more than 390 atoms to this set, resulting in $N_\text{ref}=18$.
For the assessment of the models, the atomic forces of all non-hydrogen atoms within 4.0\:\AA~of the central carbon atom were considered~($N_\text{repr}=11$).
The analysis was performed on the initial (reactant) structure of the esterification reaction and
the results are presented in Fig.~\ref{fig:mean_errors_forces_solvated_start}.

Choosing a larger radius $r_0$ and cutting probability $p$ to generate the models with a larger QM region (in contrast to 
the setting in section~\ref{sec:example_small}),
resulted in the observation that all QM/SFAM models with more than 100 QM atoms include the thiol group SH$_{5.7}$ into the QM region. We also observed that
generating the QM regions in a systematic way leads to the continuous increase of accuracy with growing QM region size.
The models taken as the reference (red points in Fig.~\ref{fig:mean_errors_forces_solvated_start}) 
do not show large fluctuations in accuracy. A mean deviation
$\varepsilon_\text{mean}^m$ of 0.05~kcal\,mol$^{-1}$\,bohr$^{-1}$ 
for this set of models was obtained
and the maximum deviation $\varepsilon_\text{max}^m$ was 0.1~kcal\,mol$^{-1}$\,bohr$^{-1}$. From this we can deduce that the functional groups that were present in some of the QM regions
of these models, but not in all of them, do not have a significant effect on the forces. Hence, we can reliably apply the reference values obtained by this strategy.
We stress that this approach is unavoidably limited by the maximum QM region size that is still computationally feasible. 
Therefore, sampling of several QM/SFAM models with QM regions
of such size is crucial in order to obtain a reliable reference.

To select a QM region from these data, we consider the models with the smallest error on the forces for a given range 
of QM region sizes (constrained by the type of calculation and the available computational resources).
To pick an example, we select a model with a QM region of $125\,\pm\,10$ atoms, which may be considered 
computationally feasible and of reasonable size for a variety of applications.
Within the data presented in Fig.~\ref{fig:mean_errors_forces_solvated_start},
this requirement is satisfied by 38 of the generated models. Note that one would typically generate only those models
fulfilling the desired size requirement and the number of candidates can be increased if deemed necessary and feasible (by varying $r_0$ and $p$).

We apply a tolerance of 0.05~kcal\,mol$^{-1}$\,bohr$^{-1}$ (based on the mean deviation in the reference models),
which yields seven models with highest accuracy of the forces as the remaining candidates.
We stress that our approach is not able to discriminate between these models reliably, as the differences in performance of these models are smaller than the mean deviation in the reference models.
An option to overcome this issue is to perform additional reference calculations on other structures on the PES and deploy these reference data to determine the optimal QM region. Calculating data for more than one point on the PES also facilitates the use of energy differences as a selection criterion. 

A simpler alternative is to base the final selection on heuristic rules. We chose to apply the following: (i) models with fewer cuts at covalent bonds (i.e., a smaller number of link atoms $m_\text{link}$) are always preferred within this pre-selection and
(ii) models with symmetry scores $m_\text{sym}$ (see Eq.~(\ref{eq:symm_score})) that are 50\% larger than the minimal $m_\text{sym}$ are discarded to prevent an
unphysical QM region to be selected due to error compensation.
In our current implementation, we employ these rules to guide the final selection, which represents a systematic and reproducible process. However, we plan to extend this process in future work, as outlined above, based on additional reference data generation to obtain a final selection based purely on first-principles data.

\begin{figure}[H]
\begin{center}
\vspace*{-0.5cm}
\includegraphics[width=0.45\textwidth, trim={0 0.1cm 0 0cm},clip]{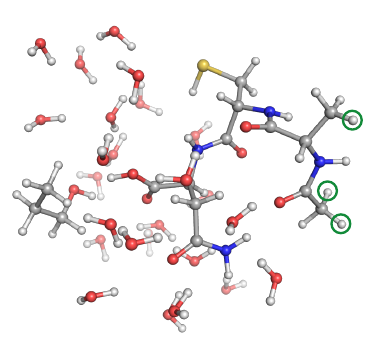}
\end{center}
\vspace*{-0.5cm}
\caption{\label{fig:selected_qm_region}\small Molecular structure of the QM region of the automatically selected QM/SFAM model. As an example, we limited the acceptable size of this QM region to $125\,\pm\,10$ atoms. The green circles indicate
the three hydrogen link atoms that are not present in the original structure, but added for valence saturation of the QM region.
Atom coloring: carbon in gray, hydrogen in white, oxygen in red, nitrogen in blue, and sulfur in yellow.}
\end{figure}

Applying this selection strategy, a QM/SFAM model with a QM region of 127 atoms (excluding 3 link atoms) was selected in our example. The molecular structure of this
QM region is depicted in Fig.~\ref{fig:selected_qm_region}. The symmetry score of this QM region is $m_\text{sym} = 1.24$ and the reaction energy error with this model
was 1.96~kcal\,mol$^{-1}$.

Finally, we emphasize that the construction of this QM/SFAM model as well as 
of its selection over the other model candidates is fully automated in our
implementation. This includes the reference calculation
management, which is automatized and parallelized in the same way as has been implemented for the SFAM parametrization process. 
We write the information about all the necessary
reference calculations into a \texttt{MongoDB} database~\cite{mongodb}, which is subsequently processed by $n$ instances of another program carrying out the calculations and storing the
results back into the database. With this set-up, $n$-fold parallelization of the data generation is enabled and therefore many QM/SFAM model candidates can be considered and tested efficiently.
This is particularly important because most of the computing time needed by the quantum region selection algorithm can be attributed to the reference calculations.
In our set-up, each of the reference calculations (QM/SFAM models with very large quantum regions) was completed in less than one hour applying 8 cores per QM calculation.


\section{Conclusions}
\label{sec:conclusions}

In this work, we reported a new QM/MM hybrid model for atomistic simulations which features,
a system-focused force field for minimized errors.
We developed a fully automated set-up of this QM/SFAM model, which by construction (i.e., by virtue of the salient features of SFAM) is not plagued with typical limitations
of standard force fields (such as missing parameters for specific metal atoms in relevant valence states). However, 
if required, the methodology reported can be combined with such a standard force field (we implemented the general AMBER force field~(GAFF)~\cite{wang04}).
Our implementation will be available within the open-source SCINE platform~\cite{scine}. 

As a result, the cumbersome manual set-up of QM/MM models has been decisively alleviated, up to the point where
it can be driven in a fully automated way, which opens up new avenues for QM/MM approaches; e.g., (i) in interactive
approaches~\cite{haag14, vaucher16}, where operator-defined abrupt changes of focus occur, (ii) in situations of quickly changing reactive sites
because of highly mobile or volatile reactants, or (iii) in studies of complex chemical systems with varying environments
such as enzymes generated by high-throughput directed evolution.

If, during a molecular exploration, new covalent bonds are formed and then shifted to the MM region
(e.g., because the QM region is moved to a different local region of the full structure), molecular-mechanics parameters may be
missing for this new chemical environment in the classical region. However, the SFAM ansatz allows our implementation to
quickly re-parametrize this new local situation with only minimal computational effort.
Furthermore, our implementation is flexible enough to allow for two or more (unconnected) QM regions in the model.

Our automated model construction process also allows for the generation and application of several models with differently sized QM regions in parallel. This enables us to estimate and control the uncertainty of the model constantly, even in fully automated exploration set-ups~\cite{simm19, unsleber20}. As was demonstrated in section~\ref{sec:example_small}, this will be of great importance when calculating physical quantities (e.g., reaction energies) that are not directly related to the atomic forces on which we based our model selection criterion. However, the modular nature of our implementation allows for extending the selection criteria to include additional quantities if necessary.

\section*{Acknowledgments}
\label{sec:acknowledgments}
C.\,B. gratefully acknowledges support by a Kekul\'{e} Ph.D. fellowship of the Fonds der Chemischen Industrie.
The authors thank the Schweizerischer Nationalfonds for generous support (Projects 200021\_182400 to M.\,R. and 200021\_172950-1 (C.\,B.) to PD Dr. Thomas Hofstetter).


\begin{mcitethebibliography}{133}
\bibitem[Duster \latin{et~al.}(2017)Duster, Wang, Garza, Miller, and
  Lin]{duster17}
Duster,~A.~W.; Wang,~C.-H.; Garza,~C.~M.; Miller,~D.~E.; Lin,~H. {Adaptive
  quantum/molecular mechanics: What have we learned, where are we, and where do
  we go from here?} \emph{WIREs Comput. Mol. Sci.} \textbf{2017}, \emph{7},
  e1310\relax
\mciteBstWouldAddEndPuncttrue
\mciteSetBstMidEndSepPunct{\mcitedefaultmidpunct}
{\mcitedefaultendpunct}{\mcitedefaultseppunct}\relax
\EndOfBibitem
\end{mcitethebibliography}

\providecommand{\latin}[1]{#1}
\makeatletter
\providecommand{\doi}
  {\begingroup\let\do\@makeother\dospecials
  \catcode`\{=1 \catcode`\}=2 \doi@aux}
\providecommand{\doi@aux}[1]{\endgroup\texttt{#1}}
\makeatother
\providecommand*\mcitethebibliography{\thebibliography}
\csname @ifundefined\endcsname{endmcitethebibliography}
  {\let\endmcitethebibliography\endthebibliography}{}

\end{document}